\input harvmac
%\input epsf
%\draftmode

\Title{\vbox{\rightline{EFI-98-01}\rightline{hep-th/9801070}}}
{\vbox{\centerline{Probing Matrix Black Holes}}}
\vskip10pt

\baselineskip=12pt
\centerline{Miao Li \ {\it and} \ Emil Martinec} 
\medskip
\centerline{\sl Enrico Fermi Inst. and Dept. of Physics}
\centerline{\sl University of Chicago}
\centerline{\sl 5640 S. Ellis Ave., Chicago, IL 60637, USA}

\baselineskip=16pt
 
\vskip 2cm
\noindent

Black holes in matrix theory may consist of interacting clusters
(correlated domains) which saturate the uncertainty principle.
We show that this assumption qualitatively accounts for the 
thermodynamic properties of both charged and neutral black holes,
and reproduces the asymptotic geometry seen by probes.

\Date{1/98}

%%%%%%%%%%%%%%%%%%%%%%%%%%%%%%%%%%%%%%%%%%%%%%%%%%%%%%%%%%%%%%%%%%%%%%%%
%%%%%%%%%%%%%%%%%%%%%%%%%%%%%%%%%%%%%%%%%%%%%%%%%%%%%%%%%%%%%%%%%%%%%%%%
%	Emil Martinec's macros
%
%
\def\journal#1&#2(#3){\unskip, \sl #1\ \bf #2 \rm(19#3) }
\def\andjournal#1&#2(#3){\sl #1~\bf #2 \rm (19#3) }

\def\ie{{\it i.e.}}
\def\eg{{\it e.g.}}
\def\cf{{\it c.f.}}

\def\sst{\scriptscriptstyle}

\def\frac#1#2{{#1\over#2}}
\def\coeff#1#2{{\textstyle{#1\over #2}}}
\def\half{\frac12}
\def\hf{{\textstyle\half}}

\def\vev#1{\langle#1\rangle}

\def\inbar{\,\vrule height1.5ex width.4pt depth0pt}
\def\IC{\relax\hbox{$\inbar\kern-.3em{\rm C}$}}
\def\IR{\relax{\rm I\kern-.18em R}}
\def\IP{\relax{\rm I\kern-.18em P}}

%
%%%%%%%%%%%%%%%%%%%%%%%%%%%%%%%%%%%%
%
\def\np#1#2#3{Nucl. Phys. {\bf B#1} (#2) #3}
\def\npb#1#2#3{Nucl. Phys. {\bf B#1} (#2) #3}
\def\pl#1#2#3{Phys. Lett. {\bf #1B} (#2) #3}
\def\plb#1#2#3{Phys. Lett. {\bf #1B} (#2) #3}
\def\prl#1#2#3{Phys. Rev. Lett. {\bf #1} (#2) #3}

\def\prd#1#2#3{Phys. Rev. {\bf D#1} (#2) #3}

\def\cqg#1#2#3{Class. Quant. Grav. {\bf #1} (#2) #3}

\catcode`\@=11
\def\slash#1{\mathord{\mathpalette\c@ncel{#1}}}
\overfullrule=0pt

\def\underrel#1\over#2{\mathrel{\mathop{\kern\z@#1}\limits_{#2}}}

\catcode`\@=12

%%%%%%%%%%%%%%%%%%%%%%%%%%%%%%%%%%%%%%%%%%%%%%%%%%%%%%%%%%%%%%

%

\def\vev#1{\left\langle #1 \right\rangle}

\def\sh{{\rm sh}}
\def\ch{{\rm ch}}

%%%%%%%%%%%%%%%%%%%%%%%%%%%%%%%%%%%%%%%%%%%%%%%%%%%%%%%%%%%%%%
% new defs:
\def\bh{{\sst BH}}
\def\lc{{\sst LC}}
\def\pr{{\sst \rm pr}}
\def\cl{{\sst \rm cl}}

\def\xdot{{\dot x}}

\def\lpl{\ell_{{\rm pl}}}

\def\str{{\sst\rm str}}
\def\nhat{{\hat N}}
\def\bh{{\sst\rm bh}}
\def\pr{{\sst\rm pr}}
%%%%%%%%%%%%%%%%%%%%%%%%%%%%%%%%%%%%%%%%%%%%%%%%%%%%%%%%%%%%%%
%%%%%%%%%%%%%%%%%%%%%%%%%%%%%%%%%%%%%%%%%%%%%%%%%%%%%%%%%%%%%%
%	References
%
\nref\dkps{M. Douglas, D. Kabat, P. Pouliot, and S. Shenker,
hep-th/9608024; \npb{485}{1997}{85}.}
%
%\nref\joe{J. Polchinski, hep-th/9510017, \prl{75}{1995}{4724}; For a 
%review see J. Polchinski, {\it TASI Lectures on
% D-Branes,} ITP preprint NSF-ITP-96-145, hep-th/9611050.
%}
%
\nref\dps{M.R. Douglas, J. Polchinski and A. Strominger, hep-th/9703031.}
\nref\limart{M. Li, E. Martinec, hep-th/9703211; hep-th/9704134; 
hep-th/9709114.}
\nref\dvv{R. Dijkgraaf, E. Verlinde, and H. Verlinde, hep-th/9704018.}
\nref\jm{J. Maldacena, hep-th/9705053; hep-th/9709099.}
\nref\cht{I. Chepelev and A. Tseytlin, hep-th/9709087.}
\nref\bfss{T. Banks, W. Fischler, S. H. Shenker and L. Susskind,
hep-th/9610043; L. Susskind, hep-th/9704080.}
\nref\bfks{T. Banks, W. Fischler, I. Klebanov and L. Susskind,
hep-th/9709091.}
\nref\ks{I. Klebanov and L. Susskind, hep-th/9709108.}
\nref\eh{E. Halyo, hep-th/9709225.}
\nref\hormart{G. Horowitz and E. Martinec, hep-th/9710217.}
\nref\miao{M. Li, hep-th/9710226.}
\nref\bfksii{T. Banks, W. Fischler, I. Klebanov and L. Susskind,
hep-th/9711005.}
\nref\dhn{B. de Wit, J. Hoppe, and H. Nicolai, \npb{305}{1988}{545}.}
\nref\BHreview{J. Maldacena, hep-th/9607235, hep-th/9705078;
G. Horowitz, contribution to the Symposium
on Black Holes and Relativistic Stars
(dedicated to the memory of S. Chandrasekhar),
Chicago, IL, 14-15 Dec 1996, hep-th/9704072; A. Peet, hep-th/9712253.}
\nref\cbh{A. Strominger and C. Vafa, hep-th/9601029, \pl  {379}{1996}{99};
 C. Callan and J. Maldacena, hep-th/9602043, \np{472}{1996}{591};
G. Horowitz and A. Strominger, hep-th/9602051, \prl{77}{1996}{2368}.}
\nref\rotbh{J. Breckenridge, R. Myers, A. Peet and C. Vafa, hep-th/9602065,
\pl{391}{1997}{93};
J. Breckenridge, D. Lowe, R. Myers, A. Peet, A. Strominger 
and C. Vafa, hep-th/9603078, \pl {381}{1996}{423}.}
\nref\mstring{L. Motl, hep-th/9701025; T. Banks and N. Seiberg, 
hep-th/9702187;
R. Dijkgraaf, E. Verlinde and H. Verlinde, hep-th/9703030.}
\nref\joegary{G. Horowitz and J. Polchinski, \prd{56}{1997}{2180};
hep-th/9707170.}
\nref\bbpt{K. Becker, M. Becker, J. Polchinski and A. Tseytlin, 
hep-th/9706072.}
\nref\liuts{H. Liu and A. Tseytlin, hep-th/9712063.}
\nref\jeff{J. Harvey, hep-th/9706039; P. Kraus, hep-th/9709199.}
\nref\glf{R. Gregory and R. Laflamme, hep-th/9301052;
\prl{70}{1993}{2837}.}
\nref\kt{D. Kabat and W. Taylor, hep-th/9711078; hep-th/9712185.}
\nref\dira{M. Dine and A. Rajaraman, hep-th/9710174.}
\nref\cama{C. Callan and J. Maldacena, hep-th/9602051, \npb{472}{1996}{591};
S. Das and S. Mathur, hep-th/9606185, Nucl. Phys. B478 (1996) 561;
A. Dhar, G. Mandal, and S. Wadia, hep-th/9605234, \plb{388}{1996}{51}.}
\nref\hwadia{S. F. Hassan, S. R. Wadia, hep-th/9703163, \plb{402}{1997}
{43}; hep-th/9712213.}
\nref\leny{L. Susskind, hep-th/9307168, \prl{71}{1993}{2367};
hep-th/9308139, \prd{49}{1994}{6606}.}
%
%\nref\dooo{M. Douglas and H. Ooguri, hep-th/9710178.}
%
%\nref\sen{A. Sen, hep-th/9709220.}
%
%\nref\seiberg{N. Seiberg, hep-th/9710009.}
%
\nref\rey{S.-J. Rey, hep-th/9711081.}
%
%
%\nref\holo{G. `t Hooft, gr-qc/9310026; L. Susskind, hep-th/9409089,
%\jmp{36}{1995}{6377}.}
%
\nref\maldastrom{J. Maldacena and A. Strominger, hep-th/9609026,
\prd{55}{1997}{861}.}
\nref\emillc{E. Martinec, hep-th/9304037, \cqg{10}{1993}{L187};
hep-th/9311129;
D. Lowe, hep-th/9312107, \plb{326}{1994}{223}.}
\nref\lpstu{D. Lowe, J. Polchinski, L. Susskind, L. Thorlacius, and J. Uglum,
hep-th/9506138, \prd{52}{1995}{6997}.}
\nref\jax{J.D. Jackson, {\it Classical Electrodynamics},
2nd ed.; J. Wiley and sons (1975).}
%
%%%%%%%%%%%%%%%%%%%%%%%%%%%%%%%%%%%%%%%%%%%%%%%%%%%%%%%%%%%%%%

\newsec{Introduction}

The geometry of a spacetime, and in particular the causal structure,
is naturally probed by an ideal massless particle.  
In M theory, the only universal object of this sort 
is the supergraviton itself.
%\foot{The supergraviton 
%is not necessarily an ideal probe in this regard, 
%since its trajectory is perturbed by the coupling of intrinsic
%spin to the background geometry.
%However, at the linearized level, one can sometimes 
%find particular polarization states
%for which the spin connection terms drop out, so that the probe
%indeed maps out the light cones of the background geometry \limart.}
In matrix theory (or more generally in the dynamics of D-branes),
curved space geometry is the effective description of probe dynamics
in the abelianized, moduli space approximation; bending of the 
probe's path is an effect of the spatially dependent vacuum polarization
caused by sources \refs{\dkps,\dps}.  There can be no global horizons
in this description (at least at finite $N$)
since the starting point is nonabelian
dynamics in Minkowskian spacetime.
The idea of using D-branes to probe black holes has been explored
in \refs{\dps-\cht}.

Recently, states in matrix theory \bfss\ bearing the qualitative properties
of Schwarzschild black holes have been constructed
\refs{\bfks-\bfksii}.
One would like to see how to reconstruct the black hole geometry
from these states.  The model developed in 
\refs{\hormart,\bfksii} describes Schwarzschild black holes as 
a collection of matrix partons (the eigenvalues of the matrices)
bound together by the semiclassical dynamics of the `off-diagonal'
(unitary conjugation) degrees of freedom.  
The partons are treated as the principal dynamical objects.
Combining the uncertainty principle, the virial theorem, and mean
field theory, one finds that a bound state of $N$ partons
has the energy and entropy of a Schwarzschild black hole
with longitudinal momentum $P=N/R$:
\eqn\NeqS{\eqalign{
  E_\lc=&~M^2R/N\cr
  M\sim&~r_0^{D-3}/G_D\cr
  S\sim&~r_0^{D-2}/G_D\sim N\ .
}}
The special point $P=S/R$ puts the system at the transition
point between black holes and black strings; in other words, $N=S$
is the minimum value for which the system approximates 
a black hole.
To obtain the entropy, the partons must be treated as distinguishable
objects, as was implicit in \hormart\ and discussed in 
detail in \bfksii.  The justification is as follows:
Energetic arguments \refs{\hormart,\bfksii}
indicate that soft excitations of the unitary degrees of freedom 
are not energetically costly relative to the overall energy of
the resonant bound state; the presence of such a unitary mode
background entwines the permutation of the eigenvalues
with the wavefunction of the unitary modes, effectively
destroying any statistics symmetry among the eigenvalues themselves.
It is important for the validity of this argument that the
time scale of motion of the unitary modes is at least as
long as the crossing time $r_0^2/R$ of the eigenvalues
in the resonant bound state; otherwise we could integrate
out the unitary modes and obtain an effective dynamics for the 
eigenvalues which would treat them as identical particles.
This bound is easily seen to be satisfied if one crudely approximates
the unitary modes as a membrane of size $\sim r_0$ \bfksii
(the membrane is indeed an excitation of the unitary modes
\refs{\dhn,\bfss}).
The membrane time scale is $T_{memb}\sim r_0^{D-3}\lpl^3/G_DR$
\hormart, which exceeds the eigenvalue crossing time if $D>5$
($D=5$ appears to be marginal).
Also $E_{\lc,{\rm memb}}\sim r_0^4R/N\ll E_{\lc,\bh}$,
so these modes don't substantially affect the energetics.

When $N\sim S$, the black hole radius is close to the longitudinal
box size \refs{\bfks,\hormart}; one expects the statistical
properties to suffer from finite size effects (for instance,
the modes identified in \refs{\bfks,\ks} as responsible for
black hole entropy).
Our main result is a clearer picture of the composition
of the black hole at $N\gg S$ as a collection 
of clusters or correlated domains of matrix partons,\foot{We
will use interchangeably the phrases `cluster' and `correlated domain'
to describe the objects entering the effective dynamics; 
our model is at the moment too crude to decide how the
degrees of freedom are ordered, \eg\ whether the ordering takes
place in position space (as in a ferromagnet at its Curie point),
momentum space (as in a BCS superconductor); or in matrix
space, or (as suggested by Lorentz symmetry) 
some more exotic combination 
of all of these coordinates.}
as proposed in \refs{\hormart,\miao}.\foot{The structure 
of the interactions differs, however, from that deduced in \miao.}
In this regime the boost is sufficiently large that the finite
size corrections due to the longitudinal box should be
irrelevant.
We will show that this scenario qualitatively reproduces
the asymptotic geometry and thermodynamic properties
of both charged and neutral black holes.

In section 2, we argue for this picture of the highly boosted
black hole as a collection of interacting clusters;
the picture is essentially a boosted version
of the mean field picture outlined above, with the partons replaced
by clusters of partons.  One difference will be that processes
where clusters exchange partons will have to be included
in the analysis.  We show that the same picture can account
for the entropy of charged black holes, by boosting the parton
distribution along an internal circle.
Section three derives the asymptotic geometry that should
be seen by a supergraviton probe in the presence of a (charged) black
hole, and shows how the probe effective action is consistent
with our picture of the structure of the black hole.
All our considerations are up to factors of order unity,
to which our approximations are insensitive.
In section 4 we interpret our results in the context of
previous work on black holes in string/M-theory 
\refs{\BHreview,\cbh,\rotbh}.

Incidentally, most of what is put forth here and in 
\refs{\hormart,\miao,\bfksii}
has validity independent of the ultimate fate of matrix theory,
since it applies to black holes extremely boosted along $x_{10}$ --
independent of the size $R$ of the longitudinal box,
which could be sub-Planckian in size.
On the other hand, our approach may prove useful in exploring 
macroscopic black holes in string theory using the
matrix string construction of \mstring; the
unitary mode background ought to reduce to a matrix string background
when a transverse circle is shrunk to sub-Planckian size.
One may in this way make contact with the 
correspondence principle of \joegary.

%%%%%%%%%%%%%%%%%%%%%%%%%%%%%%%%%%%%%%%%%%%%%%%%%%%%%%%%%%%%%%%%%%%%%%

\newsec{Cluster decomposition of matrix black holes}

Suppose the black hole consists of $S$ clusters, each containing
approximately $N/S$ matrix partons.  
The characteristic property of a cluster is its coherence --
interactions that probe it on size scales larger than
the correlation length affect the whole cluster.
Thus we will treat the clusters as the basic participants
in the dynamics, rather than the individual matrix partons.
The total longitudinal momentum 
of the system is $p_-=N/R$.  The cluster `mass'
(its longitudinal momentum) is $N/SR$.  If the spread of the
wavefunction of the cluster's center of mass is $r_0$,
the cluster's transverse velocity is $SR/Nr_0$ by the
uncertainty principle.  The virial theorem implies that the
kinetic and potential energies of the clusters are of the
same order.  The kinetic energy is
\eqn\ke{\eqalign{
  E_{\rm kin}\sim&~(\# {\rm clusters})\cdot m_\cl v_\cl^2 
	\sim S\, \frac{N}{SR}\left(\frac{SR}{Nr_0}\right)^2\cr
	\sim&~ \left(\frac{S}{r_0}\right)^2 \frac RN\ ,
}}
the correct scaling since $M\sim r_0^{D-3}/G_D$ and $S\sim r_0^{D-2}/G_D$.
The leading term in the interaction energy is the Newtonian
gravitational interaction between the transverse kinetic energies
of the clusters (together with additional terms of the same
order, as required by Galilean invariance).  For $N\gg S$
one must take into account such interactions in which the
clusters exchange longitudinal momentum.
These parton exchange processes are essential for the 
restoration of locality in the longitudinal direction in
the large $N$ limit, but difficult to calculate
in matrix theory.  For the purposes of our order-of-magnitude
estimate, we will approximate all such contributions
to the interaction as having the same magnitude and phase
as that coming from zero longitudinal momentum exchange.
Thus the leading interaction is roughly
\eqn\clint{\eqalign{
  E_{\rm pot}\sim &~G_D\sum_{\delta p_{-}=0}^{N/S}\sum_{a,b=1}^{S}
	f(\delta p_-)
	\frac{(m_\cl v_\cl^2)_a(m_\cl v_\cl^2)_b}{R r_{ab}^{D-4}}\cr
	\sim &~ G_D \frac NS S^2\frac{S^2R}{N^2r_0^D}\cr
	\sim &~ E_{kin}\frac{G_D S}{r_0^{D-2}}\ .
}}
Thus the virial theorem is satisfied given the proper scaling
of the entropy.  

The matrix theory effective action is expected 
to contain an infinite series of terms of the form
$(Nv^2)^{\ell+1}/r^{\ell(D-4)}$, as might be expected
from the expansion of the Born-Infeld action for zerobranes
\refs{\cht,\bbpt,\liuts} (see below).
At the transition point $N=S$, all terms of this form 
are comparable in magnitude due to the uncertainty relation
$v\sim R/r_0$ \refs{\miao,\bfksii}.  
Since the system at $N\gg S$ is, apart from
finite size effects, simply a boosting of the system at
$N=S$, the same should hold true for the ensemble of clusters.
For example, consider the spin-orbit term \jeff.
It contains a contribution which is the spin-orbit energy
of a cluster, interacting with the kinetic energy
of another cluster
\eqn\spinorb{
  V_{\rm spin-orb}\sim G_D\sum_{\delta p_-} \sum_{a,b} f(\delta p_-)
	\left(\frac{(r^{[i}p^{j]})\Sigma_{ij}}{m_\cl}\right)_a
	(m_\cl v_\cl^2)_b \frac{1}{Rr_{ab}^{D-2}}\ .
}
Evaluating this in the same way as \clint, one finds that these
two contributions to the potential are of the same order.
It is trivial to see that the other spin-orbit terms required
by Galilean invariance are comparable.  In fact, replacing
partons by clusters in the scaling analysis leads to the conclusion
that all terms in the expansion of the Born-Infeld action
are of comparable magnitude in the black hole.
In reaching this conclusion, it is important to realize that
all factors of $1/R$ in the effective action
of an individual D0-brane
are to be replaced by cluster longitudinal momenta
$m_\cl$, except for the one factor of $1/R$ which
arises from averaging the spatial Green's function over the
longitudinal direction.

This picture of the black hole as composed of (clusters of) partons 
interacting through effective forces due to the off-diagonal
degrees of freedom of the matrices is not completely accurate.
As mentioned in the introduction, there are `membrane' modes
that couple to the parton clusters whose time and energy scales
make them relevant to the dynamics.  Moreover, the virial theorem is 
a crucial ingredient of the argument; yet it actually says that
the action of the fluctuations of off-diagonal modes 
(the `potential' term) and the diagonal modes (the uncertainty principle
saturated `kinetic term') are of the same order.
Thus, the loop expansion which integrates them out is 
breaking down.
Clearly the true state of affairs lies in a matrix wavefunction
for the black hole where much of the matrix is excited in
a rather complicated way.  The characterization of 
the dynamics via interacting clusters of partons is at
best an approximation that captures the rough bulk properties
of the black hole state; it cannot possibly be expected
to yield precise numerical coefficients.

The above picture of the microstates of a black hole as 
a collection of interacting parton clusters also explains
much of the structure of (singly) charged black holes.
The macroscopic properties of such holes are
\eqn\chbh{\eqalign{
  M \sim &~  G_D^{-1}r_0^{D-3}(\ch^2\gamma+\coeff{1}{D-3})\cr
  Q \sim &~  G_D^{-1}r_0^{D-3} \sh\gamma\ch\gamma\cr
  S \sim &~  G_D^{-1}r_0^{D-2} \ch\gamma\ .  
}}
As in the uncharged case treated in \hormart, 
the matrix Hamiltonian at small $N$ describes
black string states stretched across the longitudinal 
direction, and at sufficiently large $N$ describes
black holes localized in the longitudinal direction.
For fixed values of the mass and charge, the 
transition between the two takes place at the value
of longitudinal boost where the entropies
are the same \glf.  The quantities \chbh\ in the boosted
frame are of course unchanged, with $E_\lc\sim Me^{-\alpha}$
and $P\sim Me^{\alpha}$; on the other hand, the black string 
has\foot{We are grateful to H. Awata for
collaborating in the calculations that established
these relations.}
\eqn\chstr{\eqalign{
  M \sim &~ G_D^{-1} 
	Rr_{\str}^{D-4}(\ch^2\delta\ch^2\beta+\coeff{1}{D-3})\cr
  P \sim &~ G_D^{-1} 
	Rr_{\str}^{D-4}\sh\beta\ch\beta\ch^2\delta\cr
  Q \sim &~ G_D^{-1} 
	Rr_{\str}^{D-4}\sh\delta\ch\delta\ch\beta\cr
  S \sim &~ G_D^{-1} 
	Rr_{\str}^{D-3}\ch\delta\ch\beta\ .
}}
In contrast to the uncharged case, the matching of these quantities
does not uniquely specify the parameters of the black string
in terms of those of the black hole; further assumptions are
required.  Demanding that the metric coefficients match smoothly
onto one another forces the horizon radii $r_0=r_\str$, boost
rapidities $\alpha=\beta$, and charge 
parameters $\gamma=\delta$ all to be equal; the rapidity
of the longitudinal boost at the transition is 
determined by $r_0=Re^{\alpha}$, the same
as in the uncharged case.
Loosely speaking, for this value of the boost the charged
black hole `just fits inside the longitudinal box'.

It is important to note that the entropy at the transition
is related to the longitudinal boost by $P=S\ch\gamma/R$;
in other words $N=S\ch\gamma\equiv\nhat$ is greater than the number
of `bits of information' stored in the charged black hole.
This is entirely reasonable, since as one increases the charge
at fixed mass to move toward extremality, the longitudinal
boost required to maintain the validity of IMF kinematics does not
decrease, while the entropy of the hole does decrease.
Thus even at the transition point $N=\nhat$, the hole is made up of
clusters of partons, with $\nhat/S>1$ partons per cluster.

The cluster hypothesis is also consistent with the entropy
of charged black holes.
The simplest way to obtain a nonextremal charged black hole
is by boosting an uncharged black hole along an internal circle.  
Let us assume that there are $S$ clusters each containing 
of order $N/S$ partons, with transverse velocity $v_\cl\sim SR/Nr_0$
in the noncompact dimensions, and internal velocity
$w_\cl\sim (SR/Nr_0)\sh\gamma$ to account for the charge.
The cluster `mass' is again $m_\cl\sim N/SR$.
The entropy of the ensemble in the transverse rest frame of the clusters
should be that of the uncharged black hole,
$S\sim r_0^{D-2}/G_{D,\rm rest}$.  However, the circle
along which the partons are moving must have proper size $L\ch\gamma$
in order that its size be $L$ after boosting along that circle;
thus $G_{D,\rm rest}=G_D/\ch\gamma$ (where $G_D=\lpl^9/L^d$
is the usual Newton constant in $D$ dimensions; we take all the 
compactification radii to be $O(L)$ for simplicity).
Thus the entropy is $S\sim r_0^{D-2}\ch\gamma/G_D$, \cf\ \chbh.  
The kinetic energy of the clusters
in the noncompact directions is
\eqn\ncke{
	E_{\rm kin}\sim \hf\sum_\cl m_\cl v_\cl^2
	\sim \frac{S^2 R}{r_0^2 N}\ .
}
The virial theorem requires that the interaction energy of the 
clusters be of the same order as this transverse kinetic
energy in the noncompact directions.  
Remarkably, following the logic of \clint, one has 
\eqn\ncpot{\eqalign{
	E_{\rm pot}\sim &~ G_D\sum_{\delta p_-=0}^{N/\nhat}\sum_{a,b=1}^{S}
		f(\delta p_-)
		\frac{(m_\cl v_\cl^2)_a(m_\cl v_\cl^2)_b}{Rr_{ab}^{D-4}}\cr
	\sim &~ E_{\rm kin} \frac{S^2G_D}{\nhat r_0^{D-2}}\ ;
}}
as argued above, the last factor is of order one.
Note that we have assumed that the maximum longitudinal momentum
transfer between clusters is $N/\nhat$ rather than 
the larger quantity $N/S$, 
since the former is the ratio of the longitudinal box size 
to the size of the hole in the highly boosted frame
and represents the effective resolution in the
longitudinal direction.
Finally, the total energy is 
\eqn\etot{\eqalign{
  E_\lc \sim &~ E_{\rm kin}(1+\sh^2\gamma)+E_{\rm pot}\cr
	\sim &~ \frac{S^2R}{r_0^2N}[\ch^2\gamma+a]\ ,
}}
where $a$ is a number of order unity; we again find qualitative
agreement with \chbh.

\newsec{Probe dynamics}

The action of a probe zerobrane (supergraviton) in a background
gravitational field may be derived either from the constrained
Hamiltonian dynamics of a massless particle, as in \limart;
or via the massless limit of the Routhian, as in 
\bbpt.\foot{We concentrate on the bosonic terms only.}
The latter route is somewhat simpler in the present context.
Let $x^+=\tau$ be the probe time coordinate; then the Routhian
is simply $S=-p_-\xdot^-$, where $\xdot^-$ is determined from
the light cone condition $ds^2=0$
\eqn\lcone{
  0 = g_{--}(\xdot^-)^2+2g_{+-}\xdot^- + g_{++} + g_{ij}v^iv^j
	+ 2g_{i-}\xdot^-v^i + 2g_{i+}v^i\ ;
}
here $v^i$ is the probe transverse velocity.
The probe action is thus (up to a total derivative)
\eqn\praction{
  S_\pr=p_-\int d\tau \frac1{g_{--}}\biggl[(g_{+-}+g_{-i}v^i)
	-\Bigl((g_{+-}+g_{-i}v^i)^2-g_{--}(g_{++}+g_{ij}v^iv^j
		+2g_{+i}v^i)\Bigr)^{1/2}\biggr]\ .
}
The leading order large distance terms in the metric scale as 
\eqn\metriccoeff{
  g_{\mu\nu}\sim \eta_{\mu\nu} + a_{\mu\nu} 
	\left(\frac{r_0}{r}\right)^{D-3}\ .
}
Expanding \praction\ to this order gives
\eqn\expanded{
  S=p_-\int \left[\hf v^2 +\hf\left(\frac{r_0}{r}\right)^{D-3}
	\left[a_{++}+a_{ij}v^iv^j+a_{+i}v^i
		-(a_{+-}+a_{-i}v^i)v^2+\coeff14 a_{--}v^4\right]
	+\ldots\right]\ .
}
The relation between the parameter $r_0$ and the ADM mass is 
\eqn\admmass{
  r_0^{D-3}=\frac{4\pi G_D M}{(D-2)\omega_{\sst D-2}}\ ,
}
where $\omega_{\sst D-2}$ is the solid angle in $D$ spacetime dimensions.
Also we will need to sum over images along the periodically
identified longitudinal direction.  Let us denote 
\eqn\transdist{
  \rho^2=r^2-\coeff14 (x^+e^{-\alpha}+x^-e^\alpha)^2
}
as the distance along the transverse noncompact dimensions.  
We have included a boost of the source 
by a rapidity $\alpha$ in the longitudinal
direction in order to facilitate passage to the infinite momentum frame.
At long distances $\rho\gg Re^\alpha$ (where $x^-$ has period $2\pi R$),
the leading nontrivial term in the metric is just the smearing of 
\metriccoeff\ across $x^-$, leading to 
\eqn\smear{
  \frac1{r^{D-3}}\longrightarrow \frac{e^{-\alpha}}{\pi R \rho^{D-4}}
	\int_{-\infty}^\infty dt(1+t^2)^{(3-D)/2}
	= \frac{e^{-\alpha}}{R\rho^{D-4}} 
	\frac{(D-3)\omega_{{\sst D-2}}}{\pi(D-4)\omega_{{\sst D-3}}}\ .
}
All told, in the presence of a source which has been boosted 
in the compact longitudinal direction,
one finds the following probe effective action 
\eqn\finalpraction{
  S_\pr = p_-\int d\tau \biggl[ \hf{v^2} + 
	\frac{A r_0^{D-3}e^{-\alpha}}{R\rho^{D-4}}
	\biggl( a_{++}+a_{ij}v^iv^j+a_{+i}v^i
                -(a_{+-}+a_{-i}v^i)v^2+\coeff14 a_{--}v^4
	\biggr)+\ldots\biggr]\ ,
}
where $A=\frac{(D-3)\omega_{{\sst D-2}}}{2\pi(D-4)\omega_{{\sst D-3}}}$.

%%%%%%%%%%%%%%%%%%%%%%%%%%%%%%%%%%%%%%%%%%%%%%%%%%%%%%%%%%%%%%%

Consider a metric of the asymptotic form
\eqn\startmetric{
  ds^2=dx\cdot dx +\Bigl(\frac{r_0}r\Bigr)^{D-3}
	\sum_\mu c_\mu (dx^\mu)^2\ .
}
Boosting first by a rapidity $\gamma$ along the $x^9$ direction,
then by a large rapidity $\alpha$ along the $x^{10}$ direction
to pass to the infinite momentum frame, one finds
\eqn\afterboost{\eqalign{
  ds^2 = dx^+ dx^- + dx_i^2 +\left(\frac{r_0}{r}\right)^{D-3}
	\Bigl[&~ \coeff14\bigl[(c_0+c_{10})+(c_0+c_9)\sh^2\gamma\bigr]
		(e^{-2\alpha}(dx^+)^2 + e^{2\alpha}(dx^-)^2)\cr
	&~ +\coeff12\bigl[(c_0-c_{10})+(c_0+c_9)\sh^2\gamma\bigr]
		dx^+dx^-\cr
	&~ +\bigl[c_9+(c_0+c_9)\sh^2\gamma\bigr]dx_9^2 + c_i dx_i^2 \cr
	&~ +(c_0+c_9) \sh\gamma\ch\gamma \; dx_9
		(e^{-\alpha}dx^+ + e^{\alpha}dx^-)\Bigr]\ .
}}
This is the asymptotic IMF metric of a charged black hole.
Suppose the probe velocity in the $x^9$ direction is $w_\pr$,
and $\vec v_\pr$ in the noncompact directions.  
Then the probe effective action \finalpraction\ is 
(after suitably averaging over the longitudinal direction)
\eqn\chargeprobe{\eqalign{
  S_\pr = \hf p_-\int d\tau\biggl[{v_\pr^2+w_\pr^2}
	+ \Bigl(\frac{\rho_0}\rho\Bigr)^{D-4}\biggl( &
		\coeff14[(c_0+c_{10})+(c_0+c_9)\sh^2\gamma]e^{-2\alpha}\cr
		&+\hf[(c_0+c_9) \sh\gamma\ch\gamma e^{-\alpha}] w_\pr \cr
		&+[(c_9+\coeff12c_0-\coeff12c_{10})
			+\coeff32(c_0+c_9)\sh^2\gamma] w_\pr^2 \cr
		&+[(c_i+\coeff12c_0-\coeff12c_{10})
                        +\coeff12(c_0+c_9)\sh^2\gamma] v_{\pr,i}^2 \cr
		&-\hf[(c_0+c_9)\sh\gamma\ch\gamma e^\alpha] 
			w_\pr(v_\pr^2+w_\pr^2) \cr
		&+\coeff1{16}[(c_0+c_{10})+(c_0+c_9)\sh^2\gamma]
			e^{2\alpha}(v_\pr^2+w_\pr^2)^2
	\biggr) + \ldots\biggr] \ ,
}}
with $\rho_0^{D-4}\sim{r_0^{D-3}e^{-\alpha}}/{R}$ due to \smear.

To proceed further, we must choose a coordinate system; 
for illustrative purposes, let us work in Schwarzschild coordinates,
where the unboosted metric is
\eqn\schwmet{
  ds^2 = -fdt^2+f^{-1} dr^2 + r^2d\Omega_{D-2}^2 + \sum_{i=D}^9 dx_i^2\ ,
}
with $f=1-(r_0/r)^{D-3}$; in the asymptotic
metric \startmetric, $c_0=c_{r}=1$, while $c_9=c_{\sst\Omega}=0$.

We would like to compare this expectation for the probe dynamics
with the effective action of matrix theory
\eqn\smatrix{
  S_{{\sst\rm eff}} \sim \sum_a \frac {v_a^2}{R} + \frac{\lpl^9}{L^d}
	\sum_{a,b} \frac{(v_a-v_b)^4}{R^3 r^{D-4}}
}
with $N$ partons comprising the black hole and one set apart
as the probe.  The effective action, and therefore the
effective metric, seen by the probe in matrix theory 
involves an average over
the distribution of parton positions and velocities
inside the black hole.  Of course, a more precise treatment
would keep the full matrix dynamics of the source, as in \kt; 
however, the abelianized approximation
will suffice for our considerations.

The parton positions relative to the black hole center 
will only affect the subleading
terms in the metric; on the other hand, due to the direction
dependence of the velocity coupling in \smatrix, the
velocity distribution directly affects the leading term.
For example, if the transverse velocity distribution of
the partons in the black hole is isotropic,
$\vev{v^iv^j}=\frac1{D-2}\delta^{ij}\vev{v^2}$,
then the leading term in the long distance metric \metriccoeff\ is
in isotropic coordinates $a_{ij}\propto \delta_{ij}$; 
on the other hand,
if the partons are all in an S-wave spatial wavefunction, 
then the angular component of their velocity vanishes
at lowest order and one is in Schwarzschild-type coordinates 
$a_{rr}\ne 0$, $a_{\sst\Omega\Omega}=0$.

Let us assume that all the black hole parton clusters are in S-wave states
in the noncompact directions, so that $c_{\sst\Omega}=0$;
and that they have the average velocity $w_\bh$ along $x_9$.
Consider the term linear in the probe internal velocity $w_\pr$;
it is 
\eqn\wcoeff{
  \frac{r_0^{D-3}}{R\rho^{D-4}} \sh\gamma\ch\gamma \, e^{-2\alpha}
	(p_{-,\pr}w_\pr)
	\sim  \frac{G_D M e^{-\alpha}}{R\rho^{D-4}}
	\Bigl(e^{-\alpha}\frac{\sh\gamma}{\ch\gamma} \Bigr)
	(p_{-,\pr}w_\pr)
}
up to coefficients of order one.  On the other hand, the effective
interaction between the probe internal velocity and the matrix black
hole constituent clusters is
\eqn\wcoeffmat{
  G_D \frac{m_\cl(v_\bh^2+w_\bh^2)w_\bh }{R\rho^{D-4}}
	\Bigl(\frac{N_\pr}{R} w_\pr \Bigr)\ .
}
Since $m_\cl(v_\bh^2+w_\bh^2)\sim E_\lc^{\sst(\bh)}\sim M e^{-\alpha}$,
comparing with \wcoeff\ shows that
\eqn\wcluster{
  w_\bh\sim \frac{\sh\gamma}{\ch\gamma}\, e^{-\alpha}\ .
}
This is precisely the relation between boost velocity and rapidity,
with the factor of $exp[-\alpha]$ arising from the further time
dilation due to the orthogonal boost involved in passing to the infinite
momentum frame.  Note also that 
\eqn\vcluster{
  v_\bh\sim \Bigl(\frac{SR}{Nr_0}\Bigr)\sim 
	\Bigl(\frac{M}{\ch\gamma P}\Bigr)\sim 
	\frac{e^{-\alpha}}{\ch\gamma}\ ,
}
so that the total cluster kinetic energy is
\eqn\kecluster{
  ({\rm \# clusters})\cdot m_\cl(v_\bh^2+ w_\bh^2) \sim
	S\cdot\frac{N}{SR}\cdot 
	\frac{1+\sh^2\gamma}{\ch\gamma^2}e^{-2\alpha}
	\sim \frac{M^2}{P}\ ,
}
comparable to the total IMF energy of the black hole
(whereas the interaction energy is only of order 
the kinetic energy of noncompact motion; see above).

Next consider the $v_\pr^2$ and $w_\pr^2$ terms in the probe
effective action.  In the one-loop effective action of matrix theory, 
these will again come from the relevant terms 
in the $v^4/\rho^{D-4}$ interaction by averaging over the
internal motions of the black hole;
we find
\eqn\wsqcluster{
  \vev{[(v_\pr-v_\bh)^2+(w_\pr-w_\bh)^2]^2}\sim
	\ldots + 2w_\pr^2\vev{v_\bh^2+3w_\bh^2}
	       + 2v_\pr^2\vev{w_\bh^2+3v_\bh^2} +\ldots
}
The form of the $v_\pr^2$ term depends more sensitively
on assumptions about the motion of the clusters
in the noncompact directions (in \wsqcluster, we have assumed
there is only radial motion in the noncompact directions).  
The corresponding terms in the
matrix effective action to this order are
\eqn\wsqmat{\eqalign{
  ({\rm \# clusters}) \frac{G_D}{R\rho^{D-4}} &
	\Bigl[m_\cl \vev{v_\bh^2+3w_\bh^2}{m_\pr w_\pr^2}
	     +m_\cl \vev{w_\bh^2+3v_\bh^2}{m_\pr v_{\rho,\pr}^2}\Bigr] \cr
  &\sim 
   S\frac{G_D}{R\rho^{D-4}}
	\Bigl(\frac{N}{SR}\Bigr)\frac{e^{-2\alpha}}{\ch^2\gamma}
	\Bigl[{m_\pr w_\pr^2}\bigl(1+3\sh^2\gamma\bigr)
		+{m_\pr v_{\rho,\pr}^2}\bigl(3+\sh^2\gamma\bigr)\Bigr]\cr
  &\sim 
  \frac{r_0^{D-3}e^{-\alpha}}{R\rho^{D-4}}
	\Bigl[{m_\pr w_\pr^2}\bigl(1+3\sh^2\gamma\bigr)
                +{m_\pr v_{\rho,\pr}^2}\bigl(3+\sh^2\gamma\bigr)\Bigr]\ .
}}
Here we have used the approximation $r_0^{D-3}\sim G_D M/\ch^2\gamma$,
valid up to coefficients of order one.  
We wish to compare to \chargeprobe,
with $c_{\sst\Omega}=c_9=c_{10}=0$ to lowest order, and
$c_0=c_{\rho}=1$ in the chosen coordinate system.
The relevant terms in \chargeprobe\ are
\eqn\compare{
  \coeff14\frac{r_0^{D-3} e^{-\alpha}}{R\rho^{D-4}}
	\Bigl[\bigl(1+3\sh^2\gamma\bigr)p_{-,\pr}w_\pr^2 +
	\bigl(3+\sh^2\gamma\bigr)p_{-,\pr}v_{\rho,\pr}^2 \Bigr]\ ;
}
thus the cluster hypothesis
reproduces the velocity-squared terms in the probe effective action 
up to coefficients of order one.  In a similar fashion, one
finds the remaining terms as well.

The leading terms in the probe effective action are admittedly
a rather weak test of the composition of the black hole;
any collection of objects 
of the same total mass and charge
contained in a finite region
will have roughly the same properties.  It is the fact that
the cluster picture reproduces the long-distance geometry,
while at the same time explaining the entropy of both neutral
and charged black holes, that gives us confidence in its
validity.

An effect we have neglected in our analysis of the probe
dynamics are the three-body and higher interactions among
the matrix constituents; these have been discussed in \dira.
In the present context, these interactions are responsible
for the response of the probe to the gravitational
binding energy of the source parton clusters.  Since
these terms are of comparable magnitude to the probe's
response to the cluster kinetic energies, they are
expected to make at most a correction of order one to
the coefficients in the effective action \chargeprobe.

%%%%%%%%%%%%%%%%%%%%%%%%%%%%%%%%%%%%%%%%%%%%%%%%%%%%%%%%%%%%%%%%%%

\newsec{Discussion}

The picture of the black hole in matrix theory 
as a resonant bound state of clusters, 
many of whose properties resemble those of 
threshold bound states, fits rather well with the overall
view of black hole thermodynamics in string theory
\BHreview.  
Near-extremal black holes in string theory can often be
regarded as a gas of massless excitations
\refs{\cbh,\cama,\hwadia}.
For instance, in the case of 
near-extremal three- and four-charge black holes
in five and four dimensions, respectively, 
precise numerical understanding has been achieved by
representing the black hole microstates 
in terms of waves along
the mutual intersection locus of a collection
of branes \refs{\BHreview,\cama,\maldastrom}.  
The massless gas is the representation
of the near-horizon physics as seen by an asymptotic observer
who probes the black hole at long wavelengths.
The description is confined to physics outside
the horizon.  At no point does the infalling matter 
appear to lose contact with the asymptotic observer, 
nor is there a point in the description of
a probe's evolution that might conceivably represent
it hitting a singularity of the background spacetime.
In string theory limits, the corresponding D-brane/string gas 
describes the degrees of freedom on a `stretched horizon'
where infalling string matter is thermalized and 
reradiated as Hawking quanta.  
Semiclassical D-brane calculations are performed in a background
Minkowskian geometry, so that temporal and spatial intervals
are those that would be measured by an asymptotic observer.
The effects of spacetime geometry are the residue
of fluctuations of the D-brane and string degrees of freedom
\refs{\dkps,\dps,\maldastrom}.
Thus, the temperature of the D-brane gas is 
approximately the Hawking temperature
rather than some blue-shifted `local temperature' that
might be experienced by stationary observers very near the horizon
(including the gas quanta themselves).

The collection of zerobrane clusters in the above description of
highly boosted black holes is rather similar.  
The black hole thermodynamics is described by a fluid,
whose properties are neither wholly zerobrane nor wholly membrane,
but rather some of each.
The temperature of the transverse virial motion of the correlation
domains is approximately the 
(longitudinally boosted) Hawking temperature.  The size of the
system is the thermal wavelength (the horizon radius),
which is the thickness of the stretched horizon
redshifted to infinity.  
Hawking radiation is, roughly speaking, the `solar wind'
of this zerobrane/membrane (\ie\ matrix) `star'.
One might expect that, as an escaping Hawking quantum
climbs out of the gravitational well, it experiences
some redshift in its wavelength.  This cannot be more
than order one if our picture is to be self-consistent.
Thus the zerobrane cluster 
fluid would appear to be a description of the horizon physics
as measured by clocks and meter sticks at infinity.
Infalling matter would not appear to fall behind a horizon
or encounter a singularity.  One difference is that the 
description would not appear to be limited to low energies,
as may be the case in the D-brane gas \maldastrom.

Another major distinction between our picture of the matrix
black hole and the D-brane gas picture
of near-extremal black holes, is that the latter has a macroscopically
`rigid' backbone of branes (those bound together to 
form the extremal configurations); the nonextremal
excitations are then draped on this scaffolding.
This allows one to find `dilute gas' regimes where the 
density of nonextremal excitations is small, and their
interactions weak.  In the generic, highly nonextremal situation,
there is no dilute gas limit; the
kinetic and interaction energies of the constituents
are of the same order.

The following picture of the state space of matrix theory
at fixed, large $N$ emerges from our study.  The ground state
is the threshold bound state graviton of momentum $P=N/R$,
with 256 polarization states and vanishing light cone
energy $E_\lc=0$.  To achieve this, the spin and orbital
wavefunctions of the constituent partons must be highly correlated;
the spins must behave antiferromagnetically, and the zero-point
motions of bosonic and fermionic degrees of freedom must
delicately balance to zero.  In the language we have been
using to describe the black hole, we would say that the
ground state consists of a single correlated domain or cluster
of partons.  According to the relation between entropy and area,
a low-energy observer would assign Planckian dimensions to the
ground state supergraviton (for {\it any} $N$).\foot{It is not clear 
what physical significance to assign to the growth in the size
of the parton cloud in a graviton with $N$ due to zero-point motions;
it is their average position that matters, as in the membrane
example given in section 4 of \hormart, and in the mean field
analysis above.  It has not 
been necessary in our analysis to utilize any sort of 
`holographic spreading' of an object with boost \leny.
The system seems more governed by standard quantum mechanics,
and by the duality between membranes and gravitons, than by
some sort of `holographic principle'.}
As we pump energy into the system,
it becomes increasingly disordered.  Instead of being 
correlated across its entire volume, there will be separate domains
which are individually ordered much as in the ground state,
with little correlation between domains.  
As with a liquid, there is no permutation symmetry among
correlation domains, since each lives in a different
environment (of off-diagonal modes).
Ascribing a finite number of states to each domain (as in the case of the
single domain of the ground state), 
the number of domains should be of order the entropy.  
The size of these resonant bound states is
governed by the ability of the constituent domains to
resolve one another via their interactions: 
$r_0$ in the transverse directions 
due to uncertainty principle; and
$RS/N\sim e^{-\alpha}r_0$ in longitudinal direction,
again due to the Fourier resolution of the clusters.

One might wonder, what distinguishes this picture
from a collection of interacting wavepackets of gravitons
in general relativity?  For instance, $S\sim r_0^{D-2}/G_D$
gravitons of wavelength $\sim r_0$ would have roughly
the right kinetic energy, and would satisfy the virial theorem
if the static gravitational interaction were used.
The difference is that the graviton gas in
general relativity is at its Schwarzschild radius,
where it is unstable to collapse toward shorter wavelengths
(as viewed from infinity); on the other hand, in matrix
theory the clusters are stable at the scale $r_0$
because gravitational forces turn off at that scale --
gravity comes from integrating out `membrane' degrees
of freedom, an approximation which breaks down at
this point.  Including these degrees of freedom in the
dynamics stabilizes the system.
A second crucial distinction is that the new degrees of
freedom `distinguish' the clusters, whereas gravitons
always have a permutation symmetry; this allows the system
to have an enormous entropy.

We have been describing the matrix black hole
as a collection of gravitons in a diffuse, membrane-like
background.  Reversing the background and the foreground,
one might also visualize the state as a `Hagedorn phase'
of the membrane, as was proposed in the second of \limart\
in the context of six-dimensional matrix black strings.
This connects the matrix black hole to the weak-coupling
Hagedorn strings which arise through the string/black hole
correspondence principle \joegary.
The Hagedorn string has zero effective tension, so there is
no communication between different regions of the string,
just as there is effectively no communication between
domains of the Hagedorn membrane that could establish
a permutation symmetry among the domains.
In the correspondence principle, the Hagedorn string
arises when the spacetime curvature 
expected from general relativity is of order the string
scale.  The surprising feature of the
Hagedorn membrane is that it does not need a curvature
of order the Planck scale to make its appearance;
the fingering instability allows the membrane to extend
its tendrils to the weak-curvature region at the Schwarzschild radius
at little cost in energy.

%%%%%%%%%%%%%%%%%%%%%%%%%%%%%%%%%%%%%%%%%%%%%%%%%%%%%%%%%%%%%%%%%%

In order to complete the picture of black hole dynamics in
M-theory, it is important to recover the description of the
evolution experienced by freely falling observers passing
through the classical event horizon.  The horizon degrees of
freedom implicitly contain this information, spread throughout
the full matrix wavefunction of the matrix black hole in
subtle correlations.  Along the lines discussed in \limart, one would
like to carry out the matrix transformation that isolates
the probe dynamics from the geometrical background by 
carrying out the sequence of boosts that keeps it in its
proper rest frame.  A coherent macroscopic object such
as the spinning membrane \refs{\hormart,\rey} is a good
candidate for a probe -- its classical rotation acting as a proper clock,
its radius a proper measuring rod.  

Since the boost between the proper rest frame of the infalling probe and
that of asymptotic observers becomes infinite at the horizon,
it is not entirely obvious that the finite $N$ matrix theory
will allow an accurate description of classical infall.
However, in the classical limit,
the probe is kept at a fixed size relative to the black
hole, while the Planck length is taken to zero.  This limit
forces $N,R\rightarrow\infty$.
The asymptotic observer sees the clock's motion freeze
as it approaches the classical horizon.  The proper motion
has the clock execute several more ticks before its
obliteration on the `singularity'.
It would be very interesting if one could extract this
classical clock variable from the diffusion of the 
probe across the full matrix wavefunction of the resulting
black hole, and thereby reconstruct the interior geometry
from the degrees of freedom already present in the matrix description.
As we have argued before \limart, this may be the ultimate
meaning of black hole complementarity: Degrees of freedom
that describe supergravity outside the black hole
do not commute with the membrane-like degrees of freedom into which the 
probe wavefunction diffuses as it penetrates
the black hole wavefunction.  In this regard,
it is interesting that at the black hole scale $r_0$,
the off-diagonal matrix elements appear to be new degrees of freedom
not present in the low-energy description of supergravity
(where low-energy is as measured by asymptotic observers).
It is these new degrees of freedom that transform our notion
of causality in a theory of extended objects \refs{\emillc,\lpstu}.
The effective notion of causal structure is induced from
the behavior of massless probes.  Signals propagate differently
in the matrix black hole; zerobranes interact strongly
with membrane-like degrees of freedom, and there may
be no localized operational definition of causal 
structure.\foot{Alternatively, one may say that in matrix theory
the underlying causal structure is that of Minkowski space.
There is an apparent causal structure induced by
matter fluctuations which can cause complicated effects in
signal propagation (\cf\ \jax, section 7.11), but
no acausality.}
There is no separation of ingoing and outgoing null rays,
as one might have expected in weakly perturbed
general relativity.

%%%%%%%%%%%%%%%%%%%%%%%%%%%%%%%%%%%%%%%%%%%%%%%%%%%%%%%%%%%%%%%%%%%%%%

\vskip 1cm
\noindent{\bf Acknowledgments:} 
We are grateful to 
H. Awata, 
J. Harvey,
and
G. Horowitz
for discussions.
This work was supported by DOE grant DE-FG02-90ER-40560
and NSF grant PHY 91-23780.

\listrefs
\end